\documentclass[preprint,phyrev]{revtex4-2}

\usepackage[dvipdfmx]{graphicx}
\usepackage{dcolumn}
\usepackage{bm}
\usepackage{amsmath}
\usepackage{amssymb}
\usepackage{braket}

\def\lesssim{\ \raise.3ex\hbox{$<$}\kern-0.8em\lower.7ex\hbox{$\sim$}\ }
\def\gesim{\ \raise.3ex\hbox{$>$}\kern-0.8em\lower.7ex\hbox{$\sim$}\ }

\begin{document}
\begin{titlepage}
\title{Thermodynamic stability, compressibility matrix, and effects of mediated interactions in a strongly-interacting Bose-Fermi mixture}
\author{Koki Manabe}
\affiliation{Department of Physics, Keio University, Yokohama 223-8522, Japan} 
\author{Yoji Ohashi}
\affiliation{Department of Physics, Keio University, Yokohama 223-8522, Japan}

\begin{abstract}
We theoretically investigate the thermodynamic stability of a normal-state Bose-Fermi mixture, with a tunable Bose-Fermi pairing interaction $-U_{\rm BF}<0$ associated with a hetero-nuclear Feshbach resonance, as well as a weak repulsive Bose-Bose interaction $U_{\rm BB}\ge 0$. Including strong hetero-pairing fluctuations associated with the former interaction within the self-consistent $T$-matrix approximation, as well as the latter within the mean-field level, we calculate the compressibility matrix, to assess the stability of this system against density fluctuations. In the weak- and the intermediate-coupling regime with respect $-U_{\rm BF}$, we show that an effective attractive interaction between bosons mediated by density fluctuations in the Fermi component makes the system unstable below a certain temperature $T_{\rm clp}$ (leading to density collapse). When $U_{\rm BB}=0$, $T_{\rm clp}$ is always higher than the Bose-Einstein condensation (BEC) temperature $T_{\rm c}$. When $U_{\rm BB}>0$, the density collapse is suppressed,  and the BEC transition becomes possible. It is also suppressed by the formation of tightly bound Bose-Fermi molecules when the hetero-pairing interaction $-U_{\rm BF}$ is strong; however, since the system may be viewed as a molecular Fermi gas in this case, the BEC transition does not also occur.  Since quantum gases involving Bose atoms are known to be sensitive to inter-particle correlations, our results would be useful for the study of many-body properties of a Bose-Fermi mixture in a stable manner, without facing the unwanted density collapse.
\end{abstract}
\maketitle
\end{titlepage}
\section{Introduction}
\label{introduction}
Recently, Bose-Fermi mixtures have attracted much attention in cold atom physics\cite{inouye2004,ni2008,bloch2008,park2012,barbut2014,onofrio2016,macro2018}. Since one can tune the strength of a Bose-Fermi pairing interaction by adjusting the threshold energy of a hetero-nuclear Feshbach resonance\cite{chin2010}, strong-coupling properties of this gas mixture have been studied\cite{storozhenko2005,watanabe2008,fratini2010,kharga2017a,manabe2019}. In addition, tuning of an effective Bose-Bose (Fermi-Fermi) interaction mediated by Fermi (Bose) component has also been discussed\cite{heiselberg2000,bijlsma2000,desalvo2019,edri2020}. As an interesting possibility, a non-$s$-wave Fermi superfluid induced by such a boson-mediated pairing interaction has recently been proposed\cite{efremov2002,wu2016,kinnunen2018}.
\par
Bose-Fermi mixtures have also been discussed in other research fields, e.g., $^3$He-$^4$He mixture\cite{edwards1965,anderson1966,bardeen1967}, as well as a high-density QCD matter\cite{maeda2009} (where the system is regarded as a mixture of bound di-quarks (bosons) and unpaired quarks (fermions)). In condensed matter physics, as a possible route to reach high-temperature superconductivity, a nano-device consisting of a n-doped semiconductor (electron gas) immersed in an exciton-polariton BEC (bosons) has theoretically been proposed\cite{laussy2010,shelykh2010,cotlet2016}. Since a Bose-Fermi mixture in cold atom physics is simple and highly tunable, this dilute gas system is expected as a useful quantum simulator\cite{georgescu2014} for the study of these more complicated many-body quantum systems.
\par
In considering a Bose-Fermi mixture with a hetero-nuclear Feshbach resonance, besides strong-coupling effects caused by a Feshbach-induced tunable interaction, thermodynamic stability is also a crucial issue\cite{desalvo2019,modugno2002,ospelkaus2006b,ospelkaus2006a,zaccanti2006,desalvo2017,lous2018}. Indeed, the density collapse of a gas mixture of Bose and Fermi atoms into the trap center has experimentally been reported\cite{modugno2002,ospelkaus2006a,ospelkaus2006b,zaccanti2006}. To simply understand this instability, it would be helpful to recall that a single-component Bose gas with an attractive interaction is unstable\cite{Pethick,gerton2000,roberts2001,donley2001,eigen2016}. In the same manner, a Bose-Fermi mixture may also become unstable  by an effective attractive Bose-Bose interaction mediated by Fermi atoms\cite{Pethick,yu2012}. 
\par
In this paper, we theoretically investigate a Bose-Fermi mixture with a tunable Bose-Fermi pairing interaction associated with a hetero-nuclear Feshbach resonance. To include strong hetero-pairing fluctuations caused by the tunable Bose-Fermi attraction, we extend the self-consistent $T$-matrix approximation developed for two-component Fermi systems\cite{haussmann1993,haussmann1994,haussmann_text} to the case when one of the two Fermi components is replaced by bosons. We briefly note that, in cold Fermi gas physics, SCTMA has been used\cite{haussmann2007} to study the BCS (Bardeen-Cooper-Schrieffer)-BEC (Bose-Einstein condensation) crossover behavior\cite{eagles1969,leggett1980,NSR,engelbrecht1993,ohashi2002,strinat2018,ohashi2020} of $^{40}$K\cite{regal2004} and $^6$Li\cite{zwierlein2004,kinast2004,jochim2004} Fermi gases. In this paper, we employ this strong-coupling theory to evaluate the compressibility matrix of a Bose-Fermi mixture, to unifiedly examine the thermodynamic stability against density fluctuations from the weak- to strong-coupling regime. In particular, we focus on how an effective Bose-Bose attractive interaction mediated by density fluctuations of fermions makes the system unstable, and how this instability is suppressed by strong-coupling effects, as well as a direct Bose-Bose repulsion.
\par 
The stability of a Bose-Fermi mixture has been examined by many researchers by various methods: References \cite{molmer1998,viverit2000,miyakawa2000,roth2002,modugno2003} discuss this problem within the mean-field approximation. Reference \cite{shirasaki2014} goes beyond the mean-field level, to include many-body effects, although the validity is still restricted to the weak-coupling regime. The stability across a Feshbach resonance is examined by a variational method in Ref. \cite{yu2011}, which is, however, unable to treat the strong-coupling regime (where tightly bound Bose-Fermi molecules dominate over system properties). Regarding this, we emphasize that SCTMA has the advantage that it can cover from the weak- to strong-coupling regime. Of course, SCTMA also still has room for improvement. In this paper, we also assess this approach for future studies.
\par
In this paper, we assumes a uniform gas, for simplicity. We briefly note that, effects of a harmonic trap have been examined in Refs. \cite{miyakawa2000,roth2002,modugno2003}, where the critical boson number (above which the system becomes unstable) has been discussed, by treating the Bose (Fermi) component within the Gross-Pitaevskii equation (Thomas-Fermi approximation).  
\par
This paper is organized as follows: In Sec. II, we explain our formulation. In Sec. III, we discuss how an effective Bose-Bose interaction mediated by fermions makes the system unstable at various strengths of a Bose-Fermi pairing interaction. We consider effects of a direct Bose-Bose repulsion on this instability in Sec. IV, to clarify the condition for the realization of BEC  without facing the unwanted density collapse. Throughout this paper, we set $\hbar=k_{\rm B}=1$, and the system volume $V$ is taken to be unity, for simplicity.
\par
\par
\section{Formulation}
\par
We consider a gas mixture of single-component Bose atoms and single-component Fermi atoms, described by the Hamiltonian,
\begin{eqnarray}
H=\sum_{\bm{p}}
\left[
\xi_p^{\rm F}f_{\bm{p}}^\dagger f_{\bm{p}}
+
\xi_p^{\rm B}b_{\bm{p}}^\dagger b_{\bm{p}}
\right]
&-&
U_{\rm BF}\sum_{\bm{p},\bm{p'},\bm{q}}f_{\bm{p}+\frac{\bm{q}}{2}}^\dagger b_{-\bm{p}+\frac{\bm{q}}{2}}^\dagger b_{-\bm{p'}+\frac{\bm{q}}{2}}f_{\bm{p'}+\frac{\bm{q}}{2}}
\nonumber
\\
&+&
\frac{U_{\rm BB}}{2}\sum_{\bm{p},\bm{p'},\bm{q}}b_{\bm{p}+\frac{\bm{q}}{2}}^\dagger b_{-\bm{p}+\frac{\bm{q}}{2}}^\dagger b_{-\bm{p'}+\frac{\bm{q}}{2}}b_{\bm{p'}+\frac{\bm{q}}{2}}.
\label{hamitonian}
\end{eqnarray}
Here, $b_{\bm{p}}^\dagger$ ($f_{\bm{p}}^\dagger $) is the creation operator of a Bose (Fermi) atom with momentum $\bm{p}$. $\xi_p^{\alpha={\rm B,F}}={\bm p}^2/(2m_\alpha)-\mu_\alpha$ is the kinetic energy of the $\alpha$-component, measured from the chemical potential $\mu_{\alpha}$ (where $m_\alpha$ is an atomic mass). In this paper, we only deal with the mass-balanced ($m_{\rm F}=m_{\rm B}\equiv m$) and population balanced ($N_{\rm B}=N_{\rm F}\equiv N$) case, for simplicity (where $N_{\rm B}$ and $N_{\rm F}$ are the numbers of Bose and Fermi atoms, respectively). $-U_{\rm BF}(<0)$ is an attractive interaction between Bose and Fermi atoms, which is assumed to be tunable by adjusting the threshold energy of a hetero-nuclear Feshbach resonance. As usual, we measure the strength of this tunable interaction in terms of the Bose-Fermi $s$-wave scattering length $a_{\rm BF}$, given by,
\begin{equation}
\frac{4\pi a_{\rm BF}}{m}=-\frac{U_{\rm BF}}{1-U_{\rm BF}\sum_{\bm{p}}^{p_{\rm c}}\frac{m}{p^2}},
\label{aBF}
\end{equation}
where $p_{\rm c}$ is a high-momentum cutoff. 
\par
In Eq. (\ref{hamitonian}), $U_{\rm BB}(\ge 0)$ is a (direct) repulsive interaction between Bose atoms, which is nothing to do with an ``effective" Bose-Bose interaction {\it mediated by Fermi atoms} (which will appear in later discussions). We assume that it is weak and constant across a hetero-nuclear Feshbach resonance. For later convenience, we introduce another $s$-wave scattering length $a_{\rm BB}$ for $U_{\rm BB}$, given by, in the Born approximation\cite{Pethick},
\begin{equation}
\frac{4\pi a_{\rm BB}}{m}=U_{\rm BB}.
\end{equation}
Of course, the $s$-wave interaction does not work between Fermi atoms due to the Pauli's exclusion principle.
\par
Effects of $-U_{\rm BF}$ and $U_{\rm BB}$ on single-particle properties of the system can conveniently be incorporated into the self-energies $\Sigma_{\alpha={\rm B,F}}(p)$ in the Bose ($\alpha={\rm B}$) and Fermi ($\alpha={\rm F}$) single-particle thermal Green's functions,
\begin{equation}
G_{\alpha={\rm B,F}}(p)
=\frac{1}{i\omega_n^\alpha-\xi_{\bm{p}}^\alpha-\Sigma_\alpha(p)}.
\label{dressedG}
\end{equation}
Here, we have introduced the abbreviated notation, $p=(\bm{p},i\omega_n^\alpha)$, where $\omega_n^{\rm B}$ and $\omega_n^{\rm F}$ are the boson and fermion Matsubara frequencies, respectively\cite{note_omega}.
\par
To include strong hetero-pairing fluctuations associated with $-U_{\rm BF}$, we extend the self-consistent $T$-matrix approximation (SCTMA) developed in the BCS-BEC crossover physics of Fermi superfluids\cite{haussmann1993,haussmann1994,haussmann_text,haussmann2007} to the present case. An advantage of SCTMA is that it is a kind of $\Phi$-derivable approximation\cite{haussmann_text,luttinger1960,baym1961,baym1962}, which allows us to calculate thermodynamic quantities in a consistent manner.
\par
\par
\begin{figure}
\includegraphics[width=11cm]{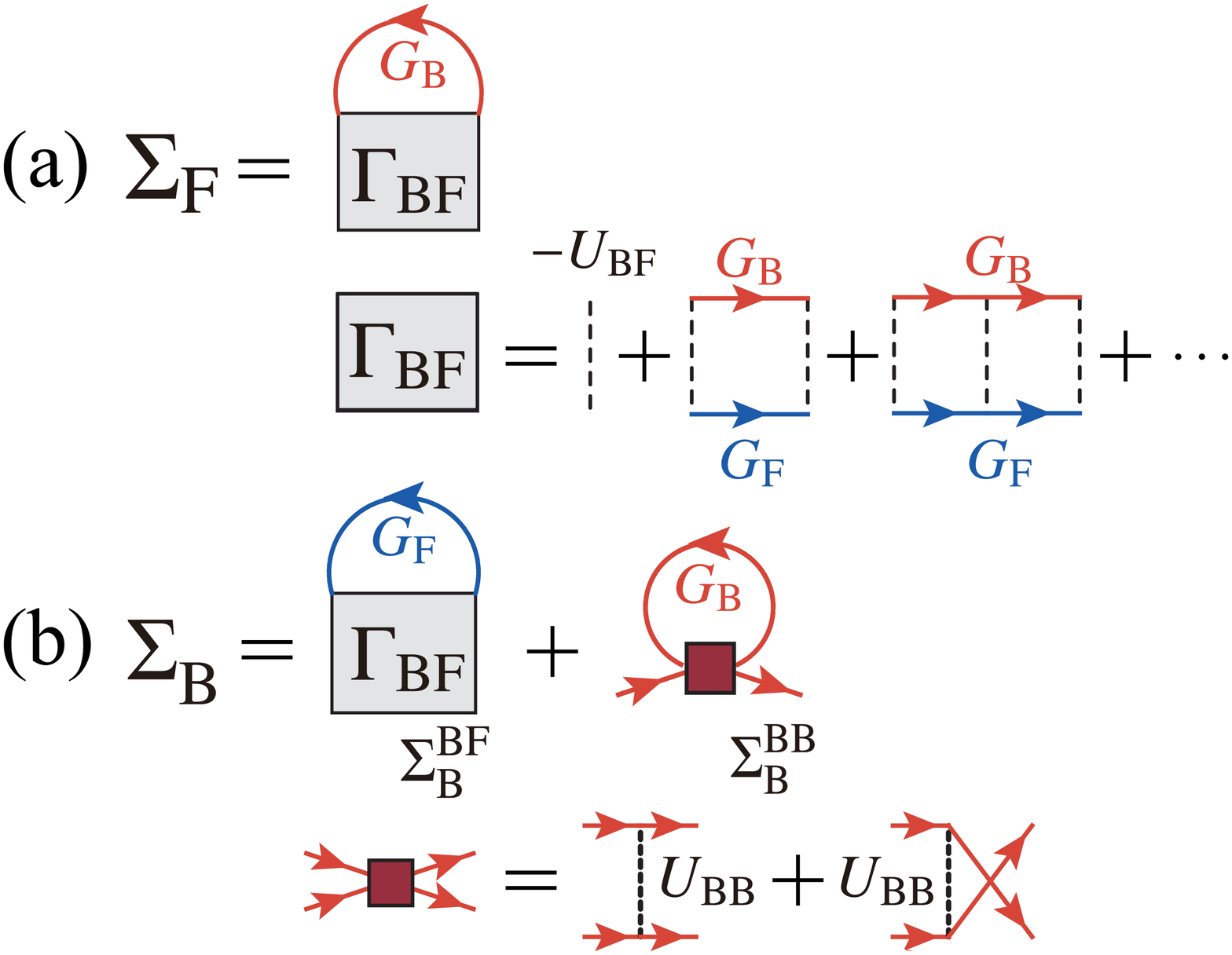}
\caption{Self-energy corrections that we are considering in this paper. (a) Fermi component $\Sigma_{\rm F}$. The particle-particle scattering matrix $\Gamma_{\rm BF}$ describes hetero-pairing fluctuations associated with the Bose-Fermi attractive interaction $-U_{\rm BF}~(<0)$ (dashed lines). $G_{\rm F}$ and $G_{\rm B}$ represent the dressed single-particle Fermi and Bose Green's function, respectively. (b) Bose component $\Sigma_{\rm B}=\Sigma_{\rm B}^{\rm BF}+\Sigma_{\rm B}^{\rm BB}$, consisting of the contribution from hetero-pairing fluctuations ($\Sigma_{\rm B}^{\rm BF}$) and that from a weak Bose-Bose repulsion $U_{\rm BB}~(\ge 0)$ ($\Sigma_{\rm B}^{\rm BB}$). For later convenience, the symmetrized interaction (red squares) is introduced for $U_{\rm BB}$. In this paper, we treat $-U_{\rm BF}$ in SCTMA, and $U_{\rm BB}$ in the mean-field level.}
\label{fig1}
\end{figure}
\par
The SCTMA fermion self-energy $\Sigma_{\rm F}$ associated with the hetero-pairing interaction $-U_{\rm BF}$ is diagrammatically drawn as Fig. \ref{fig1}(a), which gives
\begin{eqnarray}
\Sigma_{\rm F}(p)=-T\sum_q\Gamma_{\rm BF}(q)G_{\rm B}(q-p).
\label{sigmaF}
\end{eqnarray}
Here, the particle-particle scattering matrix,
\begin{eqnarray}
\Gamma_{\rm BF}(q)
&=&\frac{-U_{\rm BF}}{1-U_{\rm BF}\Pi_{\rm BF}(q)}
\nonumber
\\
&=&
\frac{1}{\frac{m}{4\pi a_{\rm BF}}+\left[\Pi_{\rm BF}(q)-\sum_{\bm p}^{p_{\rm c}}\frac{m}{p^2}\right]},
\label{Gamma}
\end{eqnarray}
physically describes hetero-pairing fluctuations. In Eq. (\ref{Gamma}),
\begin{equation}
\Pi_{\rm BF}(q)=T\sum_kG_{\rm F}(q-k)G_{\rm B}(k),
\label{eq.pi00}
\end{equation}
is the Bose-Fermi pair-correlation function.
\par
The SCTMA boson self-energy $\Sigma_{\rm B}=\Sigma_{\rm B}^{\rm BF}+\Sigma_{\rm B}^{\rm BB}$ in Fig. \ref{fig1}(b) involves effects of (1) Bose-Fermi attraction $-U_{\rm BF}$ ($=\Sigma_{\rm B}^{\rm BF}$), and (2) direct Bose-Bose repulsion $U_{\rm BB}$ ($=\Sigma_{\rm B}^{\rm BB}$). The former is given by
\begin{equation}
\Sigma_{\rm B}^{\rm BF}(p)= T\sum_q\Gamma_{\rm BF}(q)G_{\rm F}(q-p).
\label{sigmaBF}
\end{equation}
For $\Sigma_{\rm B}^{\rm BB}$, assuming that $U_{\rm BB}$ is weak, we simply treat it within the mean-field approximation, which gives
\begin{equation}
\Sigma_{\rm B}^{\rm BB}=\frac{8\pi a_{\rm BB}}{m}N.
\label{sigmaBB}
\end{equation}
\par
We briefly note that the (non-self-consistent) $T$-matrix approximation (TMA)\cite{watanabe2008,fratini2010,guidini2015} is immediately obtained by replacing all the dressed Green's functions $G_\alpha$ in Fig. \ref{fig1} with the bare ones,
\begin{equation}
G_\alpha^0(p)=\frac{1}{i\omega_n^\alpha-\xi_{\bm p}^\alpha}.
\label{TMA}
\end{equation}
As will be shown later, TMA is not suitable for our purpose, because it cannot capture the collapse of a Bose-Fermi mixture.
\par
We determine the BEC phase transition temperature $T_{\rm c}$ from the Hugenholtz-Pines theorem\cite{hugenholtz1959}, which states that the Bose excitations become gapless at $T_{\rm c}$:
\begin{equation}
\mu_{\rm B}-\Sigma_{\rm B}(p=0)=0.
\label{HP}
\end{equation}
We solve Eq. (\ref{HP}), together with the number equations,
\begin{align}
N_{\rm B}&=-T\sum_pG_{\rm B}(p), 
\label{NB}\\
N_{\rm F}&= T\sum_pG_{\rm F}(p),
\label{NF}
\end{align}
to self-consistently determine $T_{\rm c}$ and $\mu_\alpha(T_{\rm c})$, for given interaction strengths. Above $T_{\rm c}$, we only deal with the number equations (\ref{NB}) and (\ref{NF}), to determine $\mu_\alpha(T>T_{\rm c})$. 
\par
\begin{figure}
\includegraphics[width=12cm]{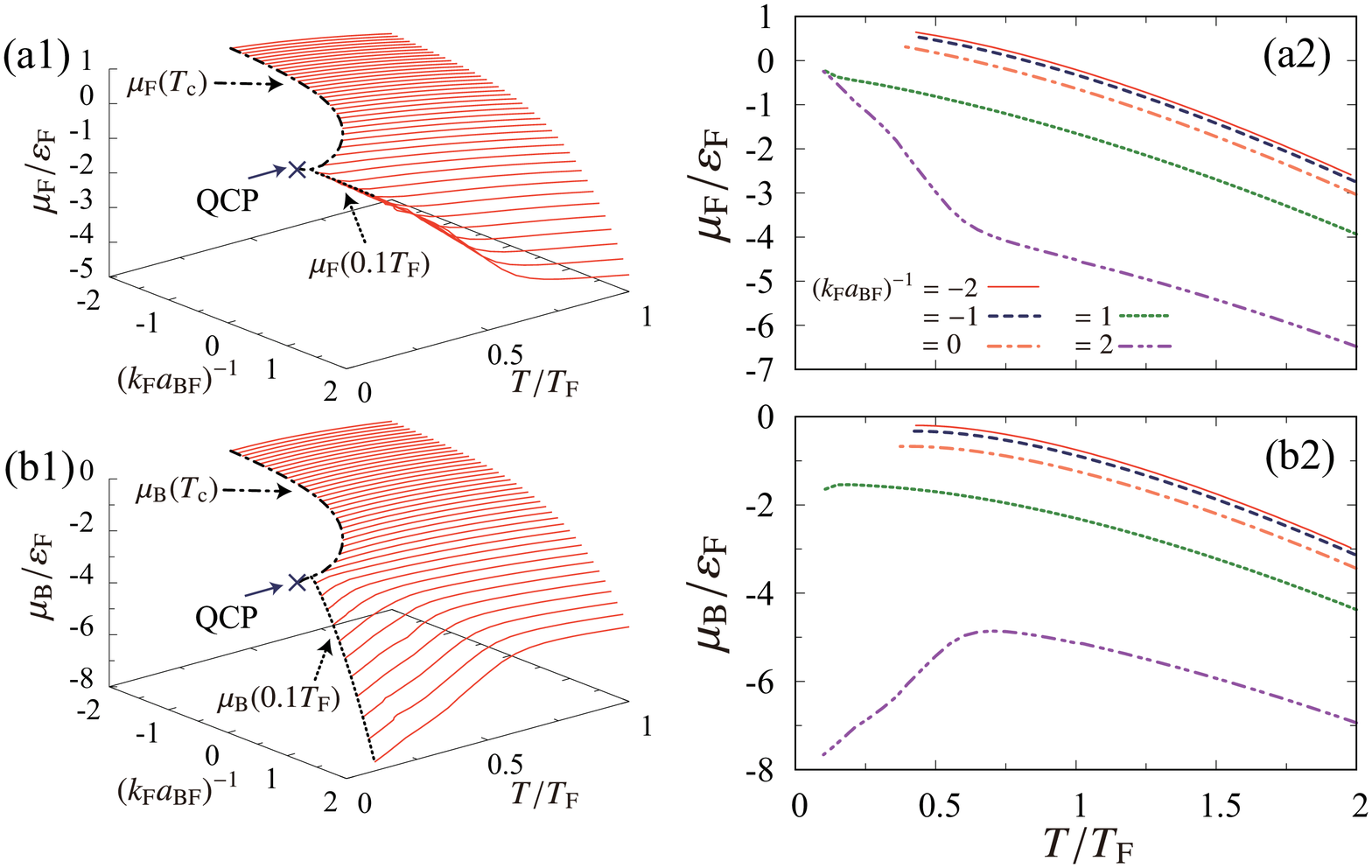}
\caption{Calculated chemical potentials $\mu_\alpha$ in SCTMA in the normal state of a Bose-Fermi mixture, above the BEC phase transition temperature $T_{\rm c}$. We set $U_{\rm BB}=0$. (a1) and (a2) $\mu_{\rm F}(T)$. (b1) and (b2) $\mu_{\rm B}(T)$. The strength of a Bose-Fermi interaction is measured in terms of the inverse scattering length $a_{\rm BF}^{-1}$ in Eq. (\ref{aBF}), normalized by the Fermi momentum $k_{\rm F}=(6\pi^2 N)^{1/3}$. $\varepsilon_{\rm F}=k_{\rm F}^2/(2m)$ and $T_{\rm F}~(=\varepsilon_{\rm F})$ are the Fermi energy and the Fermi temperature, respectively. QCP denotes the quantum critical point at which $T_{\rm c}$ vanishes. Due to numerical difficulty, calculations in the strong-coupling region are restricted to $T\ge 0.1T_{\rm F}$. We briefly note that inclusion of non-zero $U_{\rm BB}$ only causes a constant shift of $\mu_{\rm B}$ within the present mean-field approximation.}
\label{fig2}
\end{figure}
\par
Figure \ref{fig2} shows the SCTMA solutions of $\mu_{\rm F}(T\ge T_{\rm c})$ and $\mu_{\rm B}(T\ge T_{\rm c})$, when $U_{\rm BB}=0$. As previously obtained within a modified TMA scheme\cite{kharga2017a}, $T_{\rm c}$ vanishes at $(k_{\rm F}a_{\rm BF})^{-1}\simeq 0.9$ (`QCP' (quantum critical point) in this figure), and the BEC phase transition no longer occurs for stronger Bose-Fermi interactions. We briefly note that, because we treat $U_{\rm BB}$ within the mean-field approximation (see Eq. (\ref{sigmaBB})), $\mu_{\alpha={\rm F,B}}$ with $U_{\rm BB}>0$ are immediately obtained from the results in Fig. \ref{fig2} as
\begin{eqnarray}
\left\{
\begin{array}{l}
\mu_{\rm F}(U_{\rm BB}>0)=\mu_{\rm F}(U_{\rm BB}=0),\\
\mu_{\rm B}(U_{\rm BB}>0)=\mu_{\rm B}(U_{\rm BB}=0)+\Sigma_{\rm B}^{\rm BB}.
\end{array}
\right.
\label{eq.app1}
\end{eqnarray}
\par
Using the SCTMA solutions, we evaluate the compressibility matrix $\hat{\kappa}=\{\kappa_{\alpha\beta}\}$ ($\alpha,\beta={\rm F,B}$)\cite{sademelo2011}, to assess the stability of a Bose-Fermi mixture against density fluctuations. The matrix elements $\kappa_{\alpha\beta}$ are given by
\begin{equation}
\kappa_{\rm \alpha\beta}=\frac{\partial N_\alpha}{\partial \mu_\beta}.
\label{compressibility_matrix}
\end{equation}
In this paper, we numerically evaluate Eq. (\ref{compressibility_matrix}). The system is stable, if and only if $\hat{\kappa}$ is {\it positive definite}, that is, the following conditions are satisfied:
\begin{eqnarray}
\left\{
\begin{array}{l}
\kappa_{\rm BB}>0~({\rm or}~\kappa_{\rm FF}>0),\\ 
{\rm det}[\hat{\kappa}]>0.\\
\end{array}
\right. 
\label{eq.stability}
\end{eqnarray}
For the derivation of this stability condition, see Appendix A. 
\par
\begin{figure}
\includegraphics[width=14cm]{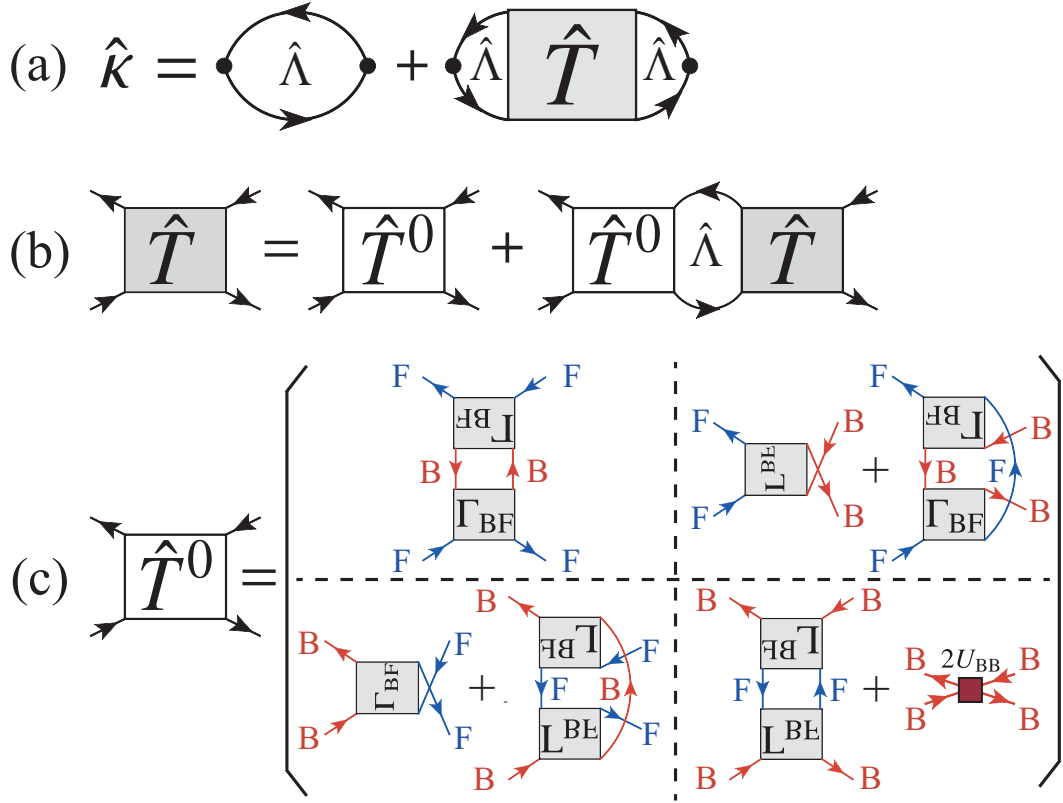}
\caption{(a) Diagrammatic representation of compressibility matrix ${\hat \kappa}$. ${\hat \Lambda}$ is given in Eq. (\ref{G2func}). ${\hat T}(p,p')$ is the $2\times 2$ matrix four-point matrix, which obeys the Bethe-Salpeter equation in (b). (c) Irreducible four-point vertex ${\hat T}^0$ appearing in (b). The particle-particle scattering matrix $\Gamma_{\rm BF}$ is given in Fig. \ref{fig1}(a). the solid lines with the label ``F" (``B") are the dressed Fermi (Bose) Green's functions in Eq. (\ref{dressedG}).}
\label{fig3}
\end{figure}
\par
We evaluate interaction corrections to the compressibility matrix ${\hat \kappa}$, so as to be consistent with the self-energy corrections $\Sigma_{\alpha={\rm F,B}}$ in Fig. \ref{fig1}. This condition is immediately satisfied, when we substitute the number equations (\ref{NB}) and (\ref{NF}) into Eq. (\ref{compressibility_matrix}). The result is
\begin{eqnarray}
\hat{\kappa}=
\left(
\begin{array}{cc}
\kappa_{\rm FF}& \kappa_{\rm FB}\\
\kappa_{\rm BF}& \kappa_{\rm BB}\\
\end{array}
\right)
=
T\sum_p\hat{\Lambda}(p)-
T^2\sum_{p,p'}\hat{\Lambda}(p)\hat{T}(p,p')\hat{\Lambda}(p'),
\label{kappa_general}
\end{eqnarray}
which is diagrammatically described as Fig. \ref{fig3}(a). Here, 
\begin{eqnarray}
\hat{\Lambda}(p)=
\left(
\begin{array}{cc}
-G_{\rm F}^2(p) & 0 \\
0& G_{\rm B}^2(p)
\end{array}
\right),
\label{G2func}
\end{eqnarray}
and the $2\times 2$ matrix four-point vertex ${\hat T}(p,p')$ obeys the Bethe-Salpeter equation\cite{Bethe}, which is diagrammatically given in Fig. \ref{fig3}(b). The expression of this equation is given by
\begin{equation}
\hat{T}(p,p')=\hat{T}^0(p,p')
-T\sum_{q}\hat{T}^0(p,q)\hat{\Lambda}(q)\hat{T}(q,p'),
\label{BSeq}
\end{equation}
where the irreducible part $\hat{T}^0$ has the form (see also Fig. \ref{fig3}(c))
\begin{eqnarray}
\hat{T}_0(p,p')
&=&
\begin{pmatrix}
0 & 0 \\
0 & \frac{8\pi a_{\rm BB}}{m}
\end{pmatrix}
+
\begin{pmatrix}
0 & \Gamma_{\rm BF}(p+p') \\
\Gamma_{\rm BF}(p+p') & 0
\end{pmatrix}
\nonumber \\
&-&
T\sum_q\Gamma_{\rm BF}^2(q)
\begin{pmatrix}
-G_{\rm B}(q-p)G_{\rm B}(q-p') &  G_{\rm B}(q-p)G_{\rm F}(q-p') \\
G_{\rm F}(q-p)G_{\rm B}(q-p') & -G_{\rm F}(q-p)G_{\rm F}(q-p')
\end{pmatrix}.
\label{T0}
\end{eqnarray}
We briefly note that the second and third term in Eq. (\ref{T0}) give, respectively, the Maki-Thompson (MT)\cite{Maki,Thompson,Kagamihara2019} and Aslamazov-Larkin (AL)\cite{Kagamihara2019,AL} type fluctuation corrections to ${\hat \kappa}$.
\par
In the non-interacting case ($U_{\rm BF}=U_{\rm BB}=0$), the first term in Eq. (\ref{kappa_general}) is reduced to the compressibility matrix in a mixture of ideal Fermi gas and ideal Bose gas, given by
\begin{eqnarray}
{\hat \kappa}_0\equiv 
\left(
\begin{array}{cc}
\kappa_{\rm FF}^0& 0\\
0&\kappa_{\rm BB}^0\\
\end{array}
\right)
=
\sum_{\bm p}
\left(
\begin{array}{cc}
\Lambda_{\rm FF}^0(\xi_{\bm p}^{\rm F})& 0\\
0& \Lambda_{\rm BB}^0(\xi_{\bm p}^{\rm B})\\
\end{array}
\right).
\label{eq.free}
\end{eqnarray}
Here, 
\begin{equation}
\Lambda_{\alpha\alpha}^0(\xi_{\bm p}^\alpha)=
{\partial f_\alpha(\xi_{\bm p}^\alpha) \over \partial\mu_\alpha},
\label{eq.freeL}
\end{equation}
where $f_{\rm F}(\xi_{\bm p}^{\rm F})$ and $f_{\rm B}(\xi_{\bm p}^{\rm B})$ are the Fermi and Bose distribution functions, respectively. As expected, Eq. (\ref{eq.free}) satisfies both the stability conditions in Eq. (\ref{eq.stability}).
\par
\begin{figure}
\includegraphics[width=15cm]{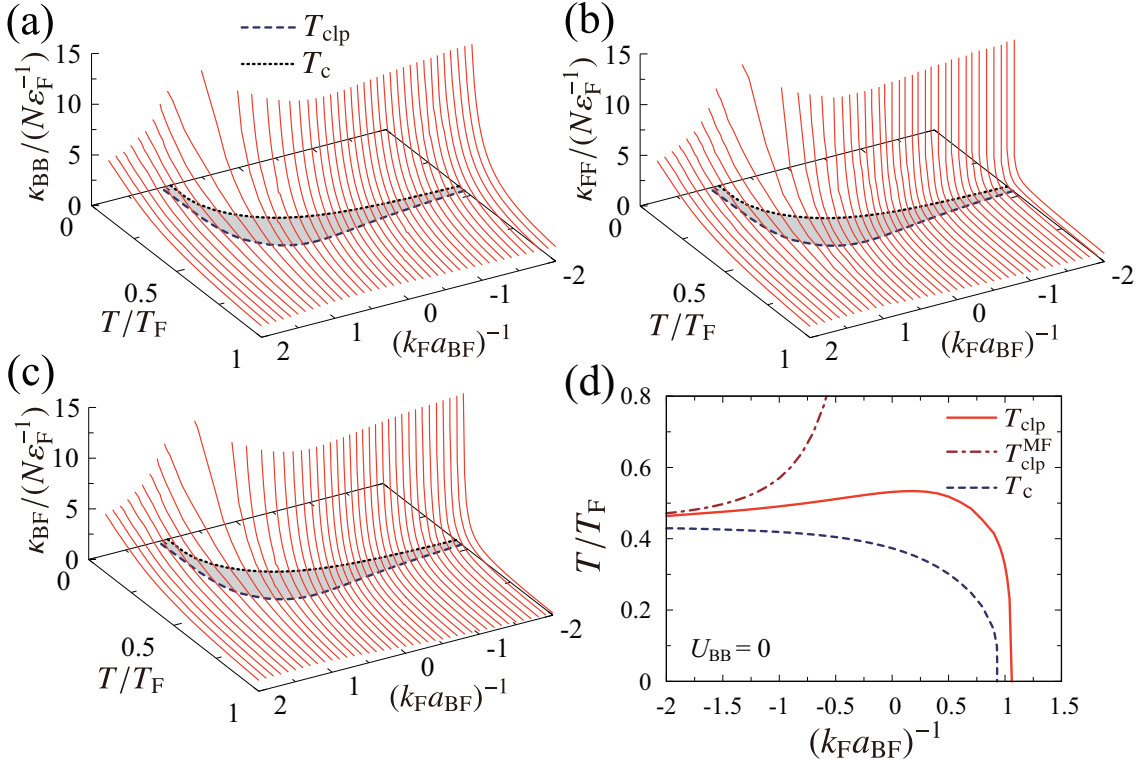}
\caption{(a)-(c) Calculated compressibilities $\kappa_{\alpha\beta}$ in a Bose-Fermi mixture, when $U_{\rm BB}=0$. (a) $\kappa_{\rm BB}$. (b) $\kappa_{\rm FF}$. (c) $\kappa_{\rm BF}~(=\kappa_{\rm FB})$. All these compressibilities diverge at the same collapse temperature $T_{\rm clp}$ shown as the dashed line in the temperature-interaction plane. In the shaded region between $T_{\rm clp}$ and the BEC phase transition temperature $T_{\rm c}$ (dotted line), the compressibilities are negative, although we do not explicitly show $\kappa_{\alpha\beta}$ there. For clarity, we show $T_{\rm clp}$ and $T_{\rm c}$ as functions of the Bose-Fermi interaction strength ($(k_{\rm F}a_{\rm BF})^{-1}$) in panel (d). $T_{\rm clp}^{\rm MF}$ is obtained from the condition that the denominator of Eq. (\ref{kappaBBRPA}) vanishes. 
}
\label{fig4}
\end{figure}

\begin{figure}
\includegraphics[width=8cm]{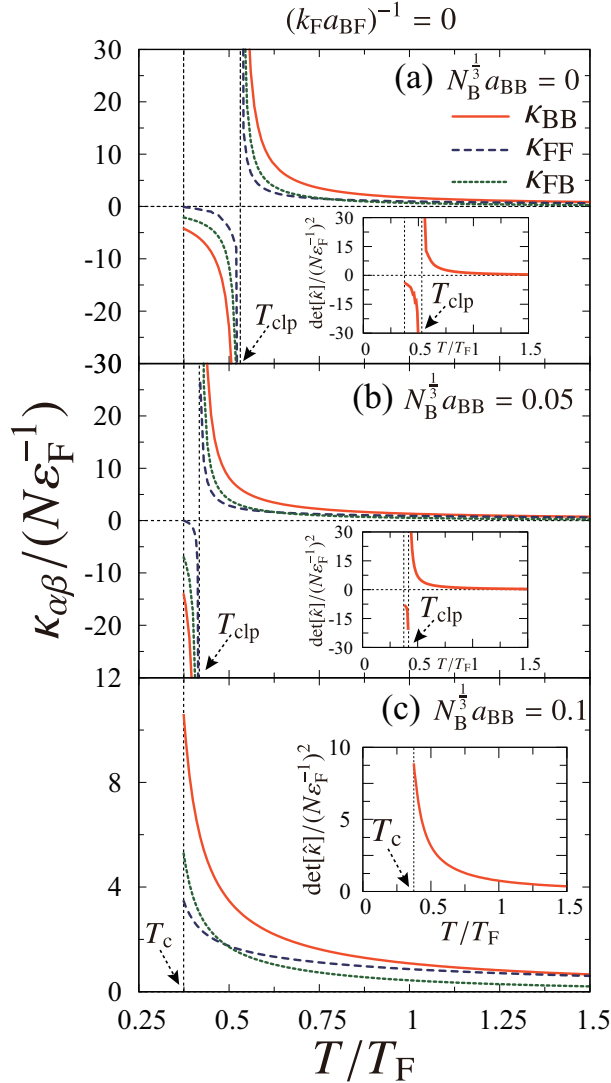}
\caption{Compressibilities $\kappa_{\alpha\beta}$ in the unitary limit ($(k_{\rm F}a_{\rm BF})^{-1}=0$). (a) $N_{\rm B}^{1/3}a_{\rm BB}=0$ ($U_{\rm BB}=0$). (b) $N_{\rm B}^{1/3}a_{\rm BB}=0.05$. (c) $N_{\rm B}^{1/3}a_{\rm BB}=0.1$. The insets in each panels show ${\rm det}[{\hat \kappa}]$. We note that $T_{\rm c}$ is unaffected by $U_{\rm BB}$ within the present mean-field approximation.}
\label{fig5}
\end{figure}
\par
\section{Stability of a Bose-Fermi mixture when $U_{\rm BB}=0$}
\par
\subsection{Weak-coupling side: Simultaneous density collapse of Bose and Fermi components}
\par
In this section, we set $U_{\rm BB}=0$. Figures \ref{fig4}(a)-(c) show the compressibilities $\kappa_{\alpha\beta}$ ($\alpha,\beta={\rm F,B}$) in this case, as functions of the temperature and the strength $(k_{\rm F}a_{\rm BF})^{-1}$ of the Bose-Fermi pairing interaction. All the compressibility components $\kappa_{\alpha\beta}$ are found to monotonically increase with decreasing the temperature, to diverge at the same collapse temperature $T_{\rm clp}$. For clarity, the interaction dependence of $T_{\rm clp}$ is separately shown in Fig. \ref{fig4}(d). This simultaneous instability of the Bose and Fermi components is consistent with the density collapse observed in $^{87}$Rb-$^{40}$K mixtures\cite{modugno2002,ospelkaus2006a,ospelkaus2006b,zaccanti2006}.
\par
To clearly show this singular behavior of $\kappa_{\alpha\beta}$ at $T_{\rm clp}$, as an example, we extract the results at the unitarity in Fig. \ref{fig5}(a): One of the stability conditions $\kappa_{\rm BB}>0$ in Eq. (\ref{eq.stability}) is found to be not satisfied below $T_{\rm clp}$, indicating the occurrence of density collapse\cite{Pethick,molmer1998,miyakawa2000,roth2002,modugno2003,viverit2000,shirasaki2014,yu2011}. Above $T_{\rm clp}$, the stability conditions in Eq. (\ref{eq.stability}) are all satisfied, so that the system is thermodynamically stable there. The same results are also obtained when $(k_{\rm F}a_{\rm BF})^{-1}\ne 0$, although we do not explicitly show the results here. We also find from Fig. \ref{fig4}(d) that, with decreasing the temperature, the system always collapses before reaching the BEC phase transition, at least when $U_{\rm BB}=0$. 
\par
To grasp the background physics of this phenomenon at $T_{\rm clp}$, it is convenient to consider the weak-coupling regime ($(k_{\rm F}a_{\rm BF})^{-1}\lesssim -1$). In this regime, since hetero-pairing fluctuations are weak,  one may safely approximate the Bose-Fermi scattering matrix $\Gamma_{\rm BF}(q)$ to the constant weak attractive interaction,
\begin{equation}
{\tilde \Gamma}_{\rm BF}\equiv {4\pi a_{\rm BF} \over m}~~~(a_{\rm BF}<0).
\label{eq.weak}
\end{equation}
The self-energies in Eqs. (\ref{sigmaF}) and (\ref{sigmaBF}) are then reduced to the mean-field ones,
\begin{equation}
\Sigma_\alpha^{\rm MF}=
{\tilde \Gamma}_{\rm BF}N_{-\alpha},
\label{eq.mf}
\end{equation}
where $-\alpha$ means the opposite component to $\alpha={\rm F,B}$. (Note that we are setting $U_{\rm BB}=0$ here, so that $\Sigma_{\rm B}=\Sigma_{\rm B}^{\rm BF}+\Sigma_{\rm B}^{\rm BB}=\Sigma_{\rm B}^{\rm BF}$.) 
\par
Using Eqs. (\ref{dressedG}), (\ref{NF}), (\ref{compressibility_matrix}), and (\ref{eq.mf}), we obtain the Bose compressibility $\kappa_{\rm BB}$ as,
\begin{eqnarray}
\kappa_{\rm BB}=
{\partial N_{\rm B} \over \partial\mu_{\rm B}}
=
T\sum_p G_{\rm B}^2(p)
\left[
1-{\partial \Sigma_{\rm B}^{\rm MF} \over \partial \mu_{\rm B}}
\right]
=
{\tilde \kappa}_{\rm BB}^0
\left[
1-{\tilde \Gamma}_{\rm BF}\kappa_{\rm FB}
\right].
\label{kappaBBMF}
\end{eqnarray}
Here, ${\tilde \kappa}_{\rm BB}^0$ is given by $\kappa_{\rm BB}^0$ in Eq. (\ref{eq.free}) with the kinetic energy $\xi_{\rm p}^{\rm B}$ being replaced by ${\tilde \xi}_{\bm p}^{\rm B}\equiv \xi_{\bm p}^{\rm B}+\Sigma_{\rm BB}^{\rm MF}$. The off-diagonal compressibility $\kappa_{\rm FB}=\partial N_{\rm F}/\partial\mu_{\rm B}$ in Eq. (\ref{kappaBBMF}) is calculated in the same manner:
\begin{eqnarray}
\kappa_{\rm FB}=
-{\tilde \kappa}_{\rm FF}^0{\tilde \Gamma}_{\rm BF}\kappa_{\rm BB},
\label{kappaFBMF}
\end{eqnarray}
where ${\tilde \kappa}_{\rm FF}^0$ is obtained from $\kappa_{\rm FF}^0$ in Eq. (\ref{eq.free}) by replacing $\xi_{\bm p}^{\rm F}$ with ${\tilde \xi}_{\bm p}^{\rm F}\equiv\xi_{\bm p}^{\rm F}+\Sigma_{\rm FF}^{\rm MF}$. Substituting Eq. (\ref{kappaFBMF}) into Eq. (\ref{kappaBBMF}), we reach
\begin{equation}
\kappa_{\rm BB}=
{{\tilde \kappa}_{\rm BB}^0 \over
1-
{\tilde \Gamma}_{\rm BF}^2{\tilde \kappa}_{\rm FF}^0{\tilde \kappa}_{\rm BB}^0
},
\label{kappaBBRPA}
\end{equation}
The Fermi compressibility $\kappa_{\rm FF}$ can also be evaluated in the same manner, giving
\begin{equation}
\kappa_{\rm FF}
={\tilde \kappa}_{\rm FF}^0+
({\tilde \Gamma}_{\rm BF}{\tilde \kappa}_{\rm FF}^0)
\kappa_{\rm BB}
({\tilde \Gamma}_{\rm BF}{\tilde \kappa}_{\rm FF}^0).
\label{kappaFFRPA}
\end{equation}
\par
In the weak-coupling regime, ${\tilde \kappa}_{\rm FF}^0$ approaches a constant value ($\simeq \rho_{\rm F}(0)>0$, where $\rho_{\rm F}(0)$ is the Fermi single-particle density of states at the Fermi level) far below the Fermi temperature $T_{\rm F}$. On the other hand, because ${\tilde \kappa}_{\rm BB}^0$ has the same form as the compressibility in an ideal Bose gas, it diverges at the BEC phase transition temperature $T_{\rm c}$. Thus, $\kappa_{\rm BB}$ in Eq. (\ref{kappaBBRPA}) always diverges at the temperature ($\equiv T_{\rm clp}^{\rm MF}>T_{\rm c}$) at which the denominator of this equation vanishes. We see in  Eq. (\ref{kappaFFRPA}) that this singularity is immediately brought about to the Fermi compressibility through the ``Bose-Fermi coupling" ${\tilde \Gamma}_{\rm BF}$, leading to the simultaneous density collapse. 
\par
We note that Eq. (\ref{kappaBBRPA}) has the same form as the compressibility in a Bose gas with the {\it attractive} interaction, 
\begin{equation}
V_{\rm BB}^{\rm eff}\equiv
-{\tilde \Gamma}_{\rm BF}^2{\tilde \kappa}_{\rm FF}^0,
\label{effectiveVBB}
\end{equation}
in the random phase approximation (RPA). Recalling that a Bose gas is unstable against an attractive interaction\cite{Pethick}, the present simultaneous collapse phenomenon is found to also come from this attractive interaction $V_{\rm BB}^{\rm eff}$. The fact that Eq. (\ref{effectiveVBB}) involves the Fermi compressibility ${\tilde \kappa}_{\rm FF}^0$ means that it is medicated by density fluctuations in the Fermi component.
\par
We compare $T_{\rm clp}^{\rm MF}$ with $T_{\rm clp}$ in Fig. \ref{fig4}(d). Although the above discussion is based on the simple approximation in Eq. (\ref{eq.weak}), the calculated $T_{\rm clp}^{\rm MF}$ agrees well with the SCTMA result in the weak-coupling regime ($(k_{\rm F}a_{\rm BF})^{-1}\lesssim -1$). 
\par
\begin{figure}
\includegraphics[width=10cm]{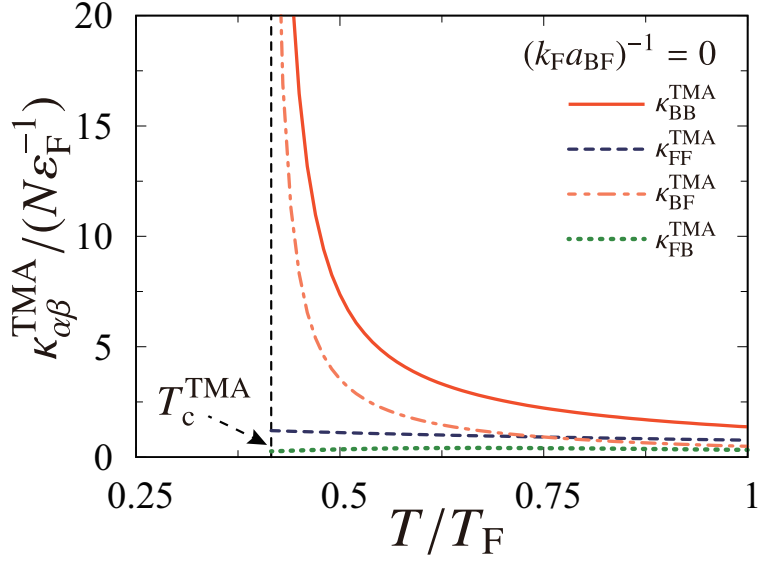}
\caption{Calculated TMA compressibilities $\kappa_{\alpha\beta}^{\rm TMA}$ in the unitary limit. $T_{\rm c}^{\rm TMA}$ is the BEC transition temperature in TMA.}
\label{fig6}
\end{figure}
\par
We emphasize that, in obtaining the above results about the simultaneous density collapse, the fact that SCTMA uses the dressed Green's functions $G_{\alpha={\rm F,B}}$ is crucial. Indeed, as shown in Fig. \ref{fig6}, this phenomenon cannot be explained in TMA, where all the dressed Green's functions in Fig. \ref{fig1} are replaced by the {\it bare} ones $G_\alpha^0$\cite{watanabe2008,fratini2010,guidini2015}: The Fermi compressibility $\kappa_{\rm FF}^{\rm TMA}$ in TMA does {\it not} diverge down to the BEC phase transition temperature. In addition, the required symmetry property,
\begin{equation}
\kappa_{\rm BF}
={\partial N_{\rm B} \over \partial \mu_{\rm F}}
=-{\partial^2 \Omega \over \partial \mu_{\rm F}\partial \mu_{\rm B}}
={\partial N_{\rm F} \over \partial \mu_{\rm B}}=
\kappa_{\rm FB},
\label{eq.symmetry}
\end{equation}
is broken in TMA, as seen in Fig. \ref{fig6}.
\par
To explain the reason for these TMA results, we again approximate $\Gamma_{\rm BF}(q)$ to ${\tilde \Gamma}_{\rm BF}$ in Eq. (\ref{eq.weak}). We then find that the key is that the particle number $N_{-\alpha}$ in Eq. (\ref{eq.mf}) (which equals $\sum_{\bm p}f_{-\alpha}(\xi_{\bm p}^{-\alpha}+\Sigma_{-\alpha}^{\rm MF})$ in SCTMA) is replaced by $N_{-\alpha}^0=\sum_{\bm p}f_{-\alpha}(\xi_{\bm p}^{-\alpha})$ in TMA, because the bare Green's functions are used in the latter theory. Noting that $\partial N_\alpha^0/\partial\mu_{-\alpha}=0$, we immediately obtain $\kappa_{\rm BB}={\tilde \kappa}_{\rm BB}^0$ and $\kappa_{\rm FF}={\tilde \kappa}_{\rm FF}^0$. That is, although the Bose compressibility $\kappa_{\rm BB}$ still diverges at the BEC phase transition (because ${\tilde \kappa}_{\rm BB}^0\to\infty$), it does not affect the Fermi compressibility $\kappa_{\rm FF}$ in the TMA case. In addition, when $N_{-\alpha}^0$ is used for $N_{-\alpha}$ in Eq. (\ref{eq.mf}), we also obtain the breakdown of the symmetry property $\kappa_{\rm BF}\ne\kappa_{\rm FB}$ as
\begin{eqnarray}
\kappa_{\rm BF}
&=&{\tilde \kappa}_{\rm BB}^0{\tilde \Gamma}_{\rm BF}\kappa_{\rm FF}^0,
\label{kappaTMABF}
\\
\kappa_{\rm FB}
&=&\kappa_{\rm BB}^0{\tilde \Gamma}_{\rm BF}{\tilde \kappa}_{\rm FF}^0\ne\kappa_{\rm BF}.
\label{kappaTMAFB}
\end{eqnarray}
\par
\begin{figure}
\includegraphics[width=10cm]{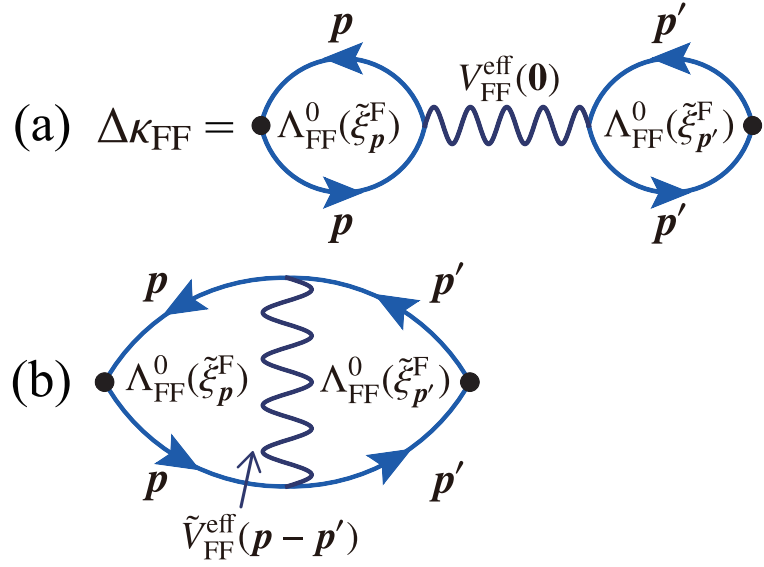}
\caption{(a) Diagrammatic representation of the last term in Eq. (\ref{kappaFFRPA}) ($=\Delta\kappa_{\rm FF}$) using the effective Fermi-Fermi interaction $V_{\rm FF}^{\rm eff}({\bm q})$ (wavy line) in Eq. (\ref{eq.VeffFF}). The solid line is the Fermi single-particle Green's function $G_{\rm F}$ with the self-energy $\Sigma_{\rm F}^{\rm MF}$ in Eq. (\ref{eq.mf}). (b) Another correction to $\kappa_{\rm FF}$. It recovers the Pauli's exclusion principle when ${\tilde V}_{\rm FF}^{\rm eff}({\bm p}-{\bm p}')=V_{\rm FF}^{\rm eff}({\bm p}-{\bm p}')$.}
\label{fig7}
\end{figure}
\par
However, the present SCTMA approach also has room for improvement, e.g., with respect to the Pauli's exclusion principle: To explain this, we again use the approximation in Eq. (\ref{eq.weak}), to rewrite the last term in Eq. (\ref{kappaFFRPA}) ($\equiv\Delta\kappa_{\rm FF}$) into the form of the first-order perturbation in terms of the effective Fermi-Fermi interaction,
\begin{equation}
H_{\rm eff}=
{1 \over 2}
\sum_{{\bm p},{\bm p}',{\bm q}}
V_{\rm FF}^{\rm eff}({\bm q})
f_{{\bm p}+{\bm q}}^\dagger
f_{{\bm p}'-{\bm q}}^\dagger
f_{{\bm p}'}
f_{\bm p},
\label{eq.effectiveFF}
\end{equation}
as
\begin{equation}
\Delta\kappa_{\rm FF}=
-\sum_{{\bm p},{\bm p}'}
\Lambda_{\rm FF}^0({\tilde \xi}_{\bm p}^{\rm F})
V_{\rm FF}^{\rm eff}(0)
\Lambda_{\rm FF}^0({\tilde \xi}_{{\bm p}'}^{\rm F}).
\label{deltakappaFF}
\end{equation}
Here, $\Lambda_{\rm FF}^0$ is given in Eq. (\ref{eq.freeL}), and 
\begin{equation}
V_{\rm FF}^{\rm eff}({\bm q})=
{
{-\tilde \Gamma}_{\rm BF}^2{\tilde \kappa}_{\rm BB}^0({\bm q})
\over
1-{\tilde \Gamma}_{\rm BF}^2
{\tilde \kappa}_{\rm FF}^0({\bm q})
{\tilde \kappa}_{\rm BB}^0({\bm q})
},
\label{eq.VeffFF}
\end{equation}
where
\begin{equation}
{\tilde \kappa}_{\alpha\alpha}({\bm q})=
\sum_{\bm k}
{f_\alpha({\tilde \xi}_{{\bm k}+{\bm q}}^\alpha)
-f_\alpha({\tilde \xi}_{\bm k}^\alpha)
\over
{\tilde \xi}_{{\bm k}+{\bm q}}^\alpha
-
{\tilde \xi}_{\bm k}^\alpha
}.
\label{eq.kappaBBQ}
\end{equation}
Noting that Eq. (\ref{deltakappaFF}) is diagrammatically described as Fig. \ref{fig7}(a), it involves the unphysical case with ${\bm p}'={\bm p}$, which corresponds to the scattering of two Fermi atoms in the {\it same} quantum state.
\par
This serious problem is removed by taking into account the other correction to $\kappa_{\rm FF}$ given in Fig. \ref{fig7}(b). This modifies Eq. (\ref{deltakappaFF}) as
\begin{equation}
\Delta\kappa_{\rm FF}=
-\sum_{{\bm p},{\bm p}'}
\Lambda_{\rm FF}^0({\tilde \xi}_{\bm p}^{\rm F})
\left[V_{\rm FF}^{\rm eff}(0)-{\tilde V}_{\rm FF}^{\rm eff}({\bm p}-{\bm p}')\right]
\Lambda_{\rm FF}^0({\tilde \xi}_{{\bm p}'}^{\rm F}).
\label{eq.Pauli}
\end{equation}
The unwanted contribution at ${\bm p}'={\bm p}$ (which contradicts with the Pauli's exclusion principle) is now canceled out by the addition term, when ${\tilde V}_{\rm FF}^{\rm eff}({\bm p}-{\bm p}')=V_{\rm FF}^{\rm eff}({\bm p}-{\bm p}')$.
\par
Regarding this problem, the SCTMA Fermi compressibility $\kappa_{\rm FF}$ involves the contribution being similar to Fig. \ref{fig7}(b), which is obtained from the (11)-component of the irreducible four-point vertex ${\hat T}^0$ shown in Fig. \ref{fig3}(c). However, it is still insufficient to fully recover the Pauli's exclusion principle. (This situation corresponds to ${\tilde V}_{\rm FF}^{\rm eff}({\bm p}-{\bm p}')\ne V_{\rm FF}^{\rm eff}({\bm p}-{\bm p}')$ in the above approximate discussion.) Because the simultaneous density collapse is caused by the divergence of $V_{\rm FF}^{\rm eff}(0)\propto\kappa_{\rm BB}$, this incomplete cancellation at ${\bm p}'={\bm p}$ means that the diverging contribution to $\kappa_{\rm F}(T=T_{\rm clp})$ around ${\bm p}'={\bm p}$ is overestimated to some extent in SCTMA. Thus, to {\it quantitatively} discuss the density collapse in a Bose-Fermi mixture, we need to improve SCTMA, which remains as our future problem. We briefly note that a similar problem has also been discussed in condensed matte physics, see Ref. \cite{senecal_text}.
\par
\begin{figure}
\includegraphics[width=10cm]{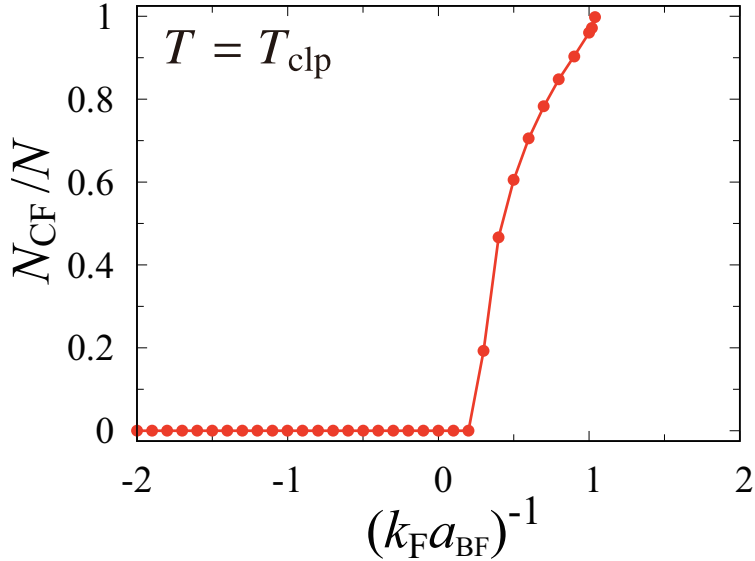}
\caption{The number $N_{\rm CF}$ of (quasi) stable Fermi molecules at $T_{\rm clp}$.}
\label{fig8}
\end{figure}
\par
\subsection{Strong-coupling side: Composite Fermi-molecular gas}
\par
We see in Fig. \ref{fig4}(d) that the collapse temperature $T_{\rm clp}$ increases with increasing the Bose-Fermi interaction strength in the weak-coupling side $(k_{\rm F}a_{\rm BF})^{-1}\lesssim 0$. Because the mean-field result $T_{\rm clp}^{\rm MF}$ also exhibits the same tendency in this regime, this behavior is considered to originate from the increase of the strength of the fermion-mediated Bose-Bose attractive interaction, as a result of the enhancement of hetero-pairing fluctuations described by $\Gamma_{\rm BF}(q)$ (see $V_{\rm BB}^{\rm eff}$ in Eq. (\ref{effectiveVBB})).
\par
However, $T_{\rm clp}$ gradually deviates from $T_{\rm clp}^{\rm MF}$, as one passes through the unitary limit, to eventually vanish at $(k_{\rm F}a_{\rm BF})^{-1}\simeq 1.1$, as shown in Fig. \ref{fig4}(d). This is due to the weakening of the bosonic character of the system, as a result of the formation of two-body composite molecular fermions in the strong-coupling side ($(k_{\rm F}a_{\rm BF})^{-1}\gesim 0$). Indeed, estimating the number $N_{\rm CF}$ of (quasi) stable molecular fermions by using the method discussed in Ref. \cite{sato2020}, we find in Fig. \ref{fig8} that $T_{\rm clp}$ vanishes, when the system becomes dominated by Fermi molecules ($N_{\rm CF}\simeq N$). (We explain the outline of how to estimate $N_{\rm CF}$ in Appendix B.) Because of the same reason, $T_{\rm c}$ also vanishes around the same interaction strength (see Fig. \ref{fig4}(d)). Thus, when $(k_{\rm F}a_{\rm BF})^{-1}\gesim 1.1$, the system may be viewed as a molecular Fermi gas, rather than an atomic Bose-Fermi mixture.
\par
\begin{figure}
\includegraphics[width=12cm]{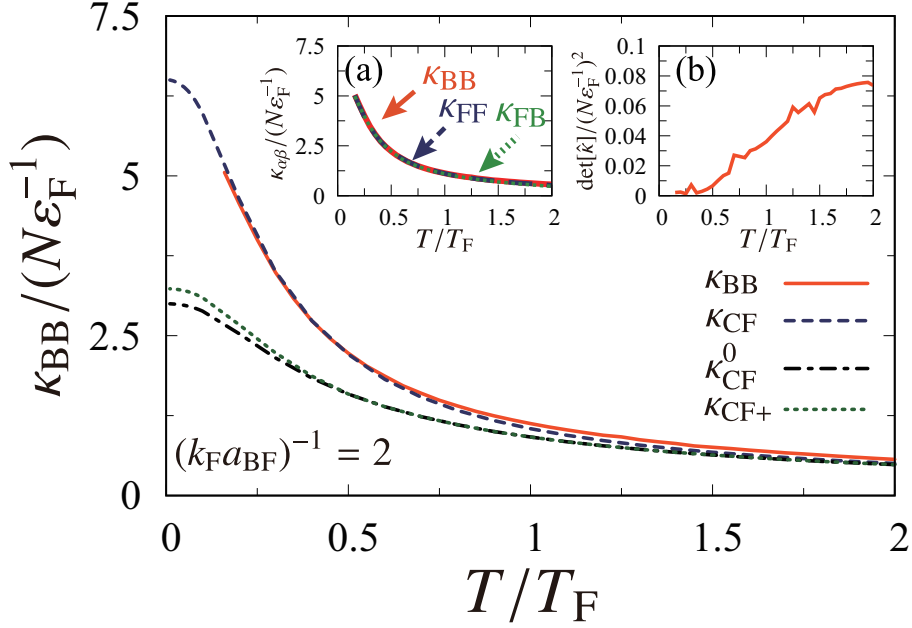}
\caption{Calculated SCTMA Bose compressibility $\kappa_{\rm BB}$ as a function of temperature, when $(k_{\rm F}a_{\rm BF})^{-1}=2$ (strong-coupling regime). $\kappa_{\rm CF}$ and $\kappa_{\rm CF}^0$ are given in Eqs. (\ref{kappasc2}) and (\ref{kappasc0}), respectively. $\kappa_{{\rm CF}+}$ shows the result in the case when the last term in Fig. \ref{fig10}(b) is added to $V_{\rm CF}^{\rm eff}$. We also show $\kappa_{\rm FF}$ and $\kappa_{\rm FB}~(=\kappa_{\rm BF})$ as functions of temperature in inset (a). Inset (b) shows ${\rm det}[{\hat \kappa}]$.}
\label{fig9}
\end{figure}
\par
Figure \ref{fig9} shows the compressibilities $\kappa_{\alpha\beta}$, when $(k_{\rm F}a_{\rm BF})^{-1}=2>1.1$ (where $T_{\rm clp}$ vanishes). In this strong-coupling case, we see in inset (a) that all the compressibility components are almost the same; however, as shown in inset (b), ${\rm det}[{\hat \kappa}]$ is still positive, at least within our numerical accuracy. Together with $\kappa_{\alpha\alpha}>0$, the system in this regime is concluded to be stable against density fluctuations. 
\par
\begin{figure}
\includegraphics[width=13cm]{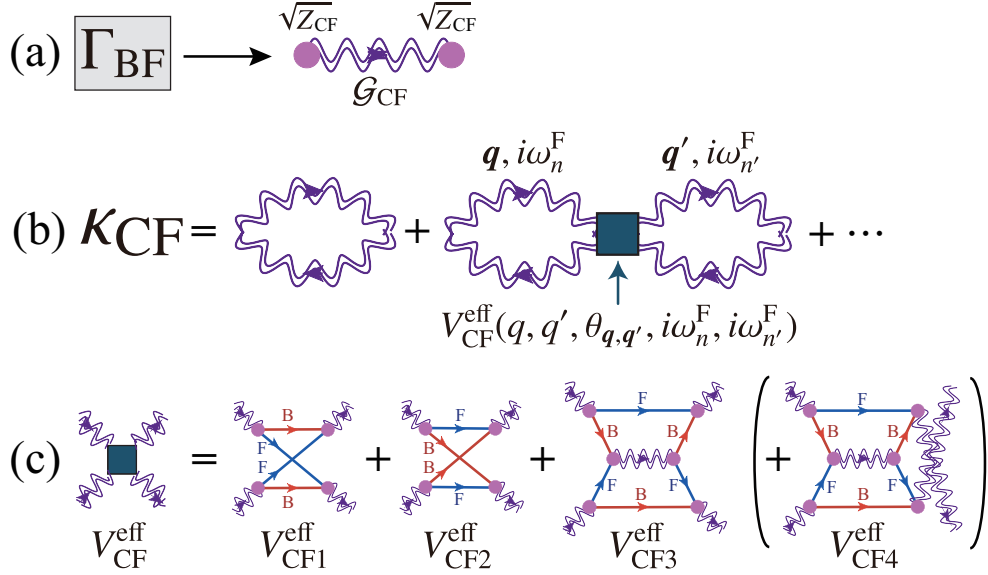}
\caption{(a) Relation between SCTMA particle-particle scattering matrix $\Gamma_{\rm BF}(q)$ and the molecular Green's function ${\mathcal G}_{\rm CF}$ in Eq. (\ref{Gcf}), deep inside the strong-coupling regime. The renormalization factor $Z_{\rm CF}$ is given below Eq. (\ref{Gamma2}). (b) Diagrammatic representation of $\kappa_{\rm CF}$ appearing in Eq. (\ref{kappasc1}). The double wavy line denotes $\mathcal{G}_{\rm CF}(q)$ in (a). (c) Effective inter-molecular interaction $V_{\rm CF}^{\rm eff}$ (filled square) mediated by unpaired atoms. While the first three terms, $V_{\rm CF1}^{\rm eff}$, $V_{\rm CF2}^{\rm eff}$, and $V_{\rm CF3}^{\rm eff}$, are involved in SCTMA, the last one, $V_{\rm CF4}^{\rm eff}$, is not. }
\label{fig10}
\end{figure}
\par
To examine how the tightly bound Bose-Fermi molecules contribute to the compressibility in Fig. \ref{fig9}, we recall that, deep inside the strong-coupling regime ($(k_{\rm F}a_{\rm BF})^{-1}\gg 1$), the particle-particle scattering matrix $\Gamma_{\rm BF}(q)$ in Eq. (\ref{Gamma}) is reduced to the molecular Green's function as\cite{haussmann1993,fratini2010} (see also Fig. \ref{fig10}(a))
\begin{equation}
\Gamma_{\rm BF}(q)\simeq Z_{\rm CF}{\mathcal G}_{\rm CF}(q),
\label{Gamma2}
\end{equation}
where $Z_{\rm CF}=8\pi/(m^2a_{\rm BF})$ and
\begin{equation}
{\mathcal G}_{\rm CF}(q)=
{1 \over i\omega_n^{\rm F}-\xi_{\bm{q}}^{\rm CF}}.
\label{Gcf}
\end{equation}
In Eq. (\ref{Gcf}), $\xi_{\bm{q}}^{\rm CF}=\bm{q}^2/(4m)-\mu_{\rm CF}$ is the molecular kinetic energy, measured from the molecular chemical potential,
\begin{equation}
\mu_{\rm CF}=\mu_{\rm F}+\mu_{\rm B}+E_{\rm b}.
\label{Gamsc}
\end{equation} 
Here, $E_{\rm b}=1/(ma_{\rm BF}^2)$ is the binding energy of a two-body Bose-Fermi bound state. In this regime ($(k_{\rm F}a_{\rm BF})^{-1}\gg 1$), the compressibility matrix ${\hat \kappa}$ in Eq. (\ref{kappa_general}) is parametrized as
\begin{equation}
\hat{\kappa}=\kappa_{\rm CF}
\left(
\begin{array}{cc}
1 & 1 \\
1 & 1
\end{array}
\right),
\label{kappasc1}
\end{equation}
where $\kappa_{\rm CF}$ is diagrammatically given in Fig. \ref{fig10}(b).  Equation (\ref{kappasc1}) is consistent with the inset (a) in Fig. \ref{fig9}, showing that all the compressibilities $\kappa_{\alpha\beta}$ take almost the same value. We briefly note that a similar ``molecular mapping" has also been discussed in the BEC regime of a two-component Fermi gas (where composite molecules are bosons)\cite{sato2020,strinati2002}. 
\par
In Fig. \ref{fig10}(b), $V_{\rm CF}^{\rm eff}$ is an effective inter-molecular interaction mediated by virtually dissociated Fermi and Bose atoms, as shown in Fig. \ref{fig10}(c). Similar diagrams have also been discussed as the origin of an effective interaction between Cooper-pairs in the BEC regime of a two-component Fermi gas\cite{haussmann1993,sato2020,pieri2000,pini2019,strinati2002}. In this paper, to analytically sum up the diagrams in Fig. \ref{fig10}(b), we approximate $V_{\rm CF}^{\rm eff}(q,q',\theta_{{\bm q},{\bm q}'},i\omega_n^{\rm F},i\omega_{n'}^{\rm F})$ in this figure to
\begin{equation}
\langle V_{\rm CF}^{\rm eff}\rangle
={1 \over 2}\int d\cos(\theta_{{\bm q},{\bm q}'}) 
V_{\rm CF}^{\rm eff}
(k_{\rm F}^{\rm CF},k_{\rm F}^{\rm CF},\theta_{{\bm q},{\bm q}'},0,0).
\label{UCFave}
\end{equation}
In Eq. (\ref{UCFave}), assuming that the region near the (molecular) Fermi surface is important\cite{abrikosov_text}, we fix $q$ and $q'$ the value at the Fermi surface $k_{\rm F}^{\rm CF}=\sqrt{4m\mu_{\rm CF}}$\cite{comment5}, and take the angular average with respect to the relative angle $\theta_{{\bm q},{\bm q}'}$ between ${\bm q}$ and ${\bm q}'$ over the Fermi surface. For Matsubara frequencies, we take the analytic continuation $i\omega_n^{\rm F},~i\omega_{\rm n'}^{\rm F}\to\omega+i\delta$ and set $\omega=0$. These approximations enable us to sum up the diagrams in Fig. \ref{fig10}(b), which gives the RPA-type expression, 
\begin{equation}
\kappa_{\rm CF}=
{
\kappa_{\rm CF}^0 \over 1+\langle V_{\rm CF}^{\rm eff} \rangle \kappa_{\rm CF}^0
},
\label{kappasc2}
\end{equation}
where 
\begin{equation}
\kappa_{\rm CF}^0=\sum_{\bm q}
{\partial f_{\rm F}(\xi_{\bm q}^{\rm CF}) 
\over 
\partial \mu_{\rm CF}},
\label{kappasc0}
\end{equation}
is the compressibility in a free molecular Fermi gas. 
\par
Evaluation of the averaged interaction $\langle V_{{\rm CF}j}^{\rm eff}\rangle$ gives\cite{comment4},
\begin{equation}
\langle V_{\rm CF}^{\rm eff}\rangle
=\sum_{j=1}^3
\langle V_{{\rm CF}j}^{\rm eff}\rangle
\simeq -\frac{4\pi\times(0.84 a_{\rm BF})}{2m}~(<0).
\label{UCFave2}
\end{equation}
Figure \ref{fig9} shows that $\kappa_{\rm CF}$ in Eq. (\ref{kappasc2}) with the averaged interaction strength $\langle V_{{\rm CF}j}^{\rm eff}\rangle$ in Eq. (\ref{UCFave2}) well describes the SCTMA result of $\kappa_{\rm BB}$.
\par
However, because the second term in Fig. \ref{fig10}(b) has the same diagrammatic structure as $\Delta\kappa_{\rm FF}$ in Fig. \ref{fig7}(a), if the $s$-wave component unphysically remains in $V_{\rm CF}^{\rm eff}$, it directly affects $\kappa_{\rm CF}$. Indeed, while the $s$-wave components of $V_{\rm CF1}^{\rm eff}$ and $V_{\rm CF2}^{\rm eff}$ are canceled out with each other as
\begin{equation}
V_{\rm CF1}^{\rm eff}|_{s{\mathchar `-}{\rm wave}}=-V_{\rm CF2}^{\rm eff}|_{s{\mathchar `-}{\rm wave}}
={4\pi a_{\rm BF} \over 2m},
\label{VCFs-wave}
\end{equation}
there is no ``counter" term to remove the $s$-wave component of $V_{\rm CF3}^{\rm eff}$ in SCTMA. In this sense, the Pauli's exclusion principle is unphysically broken in $\kappa_{\rm CF}$.
\par
To overcome this problem, we need to go beyond SCTMA, to include the fourth term $V_{\rm CF4}^{\rm eff}$ in Fig. \ref{fig10}(c) (which plays the same role as ${\tilde V}_{\rm FF}^{\rm eff}$ in Fig. \ref{fig7}(b)). Then, Eq. (\ref{UCFave2}) is replaced by, at $T=0$,
\begin{equation}
\langle V_{\rm CF}^{\rm eff}\rangle
=\sum_{j=1}^4
\langle V_{{\rm CF}j}^{\rm eff}\rangle
\simeq
-{4\pi(k_{\rm F}^{\rm CF}a_{\rm BF})^2\times (0.5a_{\rm BF}) \over 2m}~(<0).
\label{modUeff}
\end{equation}
This much weaker interaction than Eq. (\ref{UCFave2}) comes from non $s$-wave components. When this improved version is used, the compressibility becomes very close to that in the non-interacting case ($\kappa_{\rm CF}^0$), as shown in Fig. \ref{fig9}. Thus, although SCTMA can describe the stabilization of the system in the strong-coupling regime due to the formation of Bose-Fermi bound molecules, it again overestimates the magnitude of the molecular compressibility, because of the insufficient treatment of inter-molecular interaction. 
\par
\begin{figure}
\includegraphics[width=10cm]{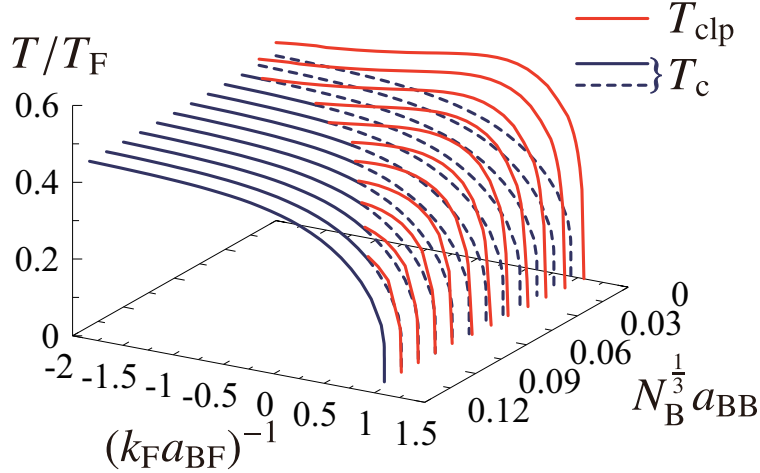}
\caption{Calculated collapse temperature $T_{\rm clp}$ in the presence of Bose-Bose repulsion $U_{\rm BB}=4\pi a_{\rm BB}/m>0$. When $T_{\rm c}<T_{\rm clp}$ we plot $T_{\rm c}$ by the dashed line.}
\label{fig11}
\end{figure}
\par
\section{Effects of Bose-Bose repulsion $U_{\rm BB}$ on thermodynamic stability}
\par
Because the density collapse discussed in the previous section comes from the effective Bose-Bose attractive interaction mediated by the Fermi component, this singular phenomenon is expected to be suppressed by the direct Bose-Bose repulsion $U_{\rm BB}=4\pi a_{\rm BB}/m>0$. Indeed, Fig. \ref{fig5} confirm this, that is, the collapse temperature $T_{\rm clp}$ decreases with increasing the interaction strength $N_{\rm BB}^{1/3}a_{\rm BB}$. In panel (c), all the compressibilities $\kappa_{\alpha\beta}$, as well as ${\rm det}[{\hat \kappa}]$, are positive everywhere above $T_{\rm c}$, indicating that the system is stabilized by the Bose-Bose repulsion $U_{\rm BB}>0$.
\par
Figure \ref{fig11} shows $T_{\rm clp}$ and effects of the Bose-Bose repulsion. One sees in this figure that, when $N_{\rm B}^{1/3}a_{\rm BB}\gesim 0.11$, one can reach the BEC phase transition without suffering from the density collapse, in the whole coupling regime with respect to the hetero-pairing interaction $(k_{\rm F}a_{\rm BF})^{-1}$.
\par
The previous discussion using Eq. (\ref{eq.weak}) is also applicable to the present case, by replacing the mean-field Bose self-energy in Eq. (\ref{eq.mf}) with
\begin{equation}
\Sigma_{\rm B}^{\rm MF}=
{\tilde \Gamma}_{\rm BF}N_{\rm F}+2U_{\rm BB}N_{\rm B}.
\label{eq.mf2}
\end{equation}
Repeating the same discussion below Eq. (\ref{eq.mf}), one reaches
\begin{eqnarray}
\hat\kappa(U_{\rm BB}>0)=
{1 \over 1
+[2U_{\rm BB}-{\tilde \Gamma}_{\rm BF}^2{\tilde \kappa}^0_{\rm FF}]
{\tilde \kappa}^0_{\rm BB}
}
\left(
\begin{array}{cc}
{\tilde \kappa}_{\rm F}^0
+2U_{\rm BB}{\tilde \kappa}_{\rm FF}^0{\tilde \kappa}_{\rm BB}^0  
&
-{\tilde \Gamma}_{\rm BF}{\tilde \kappa}_{\rm FF}^0{\tilde \kappa}_{\rm BB}^0 
\\
-{\tilde \Gamma}_{\rm BF}{\tilde \kappa}_{\rm FF}^0{\tilde \kappa}_{\rm BB}^0
&  
{\tilde \kappa}_{\rm BB}^0
\end{array}
\right).
\label{kappaMF2}
\end{eqnarray}
Equation (\ref{kappaMF2}) indicates that the compressibility matrix no longer diverges when
\begin{equation}
2U_{\rm BB}-{\tilde \Gamma}_{\rm BF}{\tilde \kappa_{\rm FF}^0}=
{8\pi a_{\rm BB} \over m}-
\left({4\pi a_{\rm BF} \over m}\right)^2\kappa_{\rm FF}^0 > 0,
\label{stbcond}
\end{equation}
which qualitatively explains the behavior of $\kappa_{\alpha\beta}$ in Fig.\ref{fig5}. Comparing the standard RPA expression for the compressibility with $\kappa_{\rm BB}$ in Eq. (\ref{kappaMF2}), one finds that the stability condition in Eq. (\ref{stbcond}) is equivalent to the realization of the situation that the interaction between Bose atoms, $2U_{\rm BB}-{\tilde \Gamma}_{\rm BF}{\tilde \kappa_{\rm FF}^0}$, is repulsive.
\par
While the stabilization of the Bose component is accompanied by the sign change of the Bose-Bose interaction $2U_{\rm BB}-{\tilde \Gamma}_{\rm BF}{\tilde \kappa_{\rm FF}^0}$, the Fermi component becomes stable somehow in a different manner: The (11)-component of Eq. (\ref{kappaMF2}) can be written as,
\begin{equation}
\kappa_{\rm FF}(U_{\rm BB}>0)=
{
{\tilde \kappa}_{\rm FF}^0
\over
1-
{\tilde \Gamma}_{\rm BF}^2\kappa^{\rm RPA}_{\rm BB}{\tilde \kappa}_{\rm FF}^0},
\label{kappaFFUB}
\end{equation}
where
\begin{equation}
\kappa^{\rm RPA}_{\rm BB}=
{
{\tilde \kappa}_{\rm B}^0
\over
1+2U_{\rm BB}{\tilde \kappa}_{\rm B}^0
}.
\label{kappaBBUBB}
\end{equation}
The RPA-type structure in Eq. (\ref{kappaFFUB}) shows that the effective interaction $-{\tilde \Gamma}_{\rm BF}^2\kappa_{\rm BB}^{\rm RPA}$ between Fermi atoms is always {\it attractive}, irrespective of the magnitude of $U_{\rm BB}\ge 0$. Since ${\tilde \kappa}_{\rm BB}^0$ monotonically increases with decreasing the temperature to diverge at the BEC phase transition, the maximum value of $\kappa_{\rm BB}^{\rm RPA}$ in Eq. (\ref{kappaBBUBB}) equals $1/(2U_{\rm BB})$. Thus, even when the stability condition in Eq. (\ref{stbcond}) is satisfied, the Fermi-Fermi interaction, $-{\tilde \Gamma}_{\rm BF}^2\kappa_{\rm BB}^{\rm RPA}$, is still attractive, but is not strong enough to cause the density collapse of the Fermi component.
\par
\begin{figure}
\includegraphics[width=10cm]{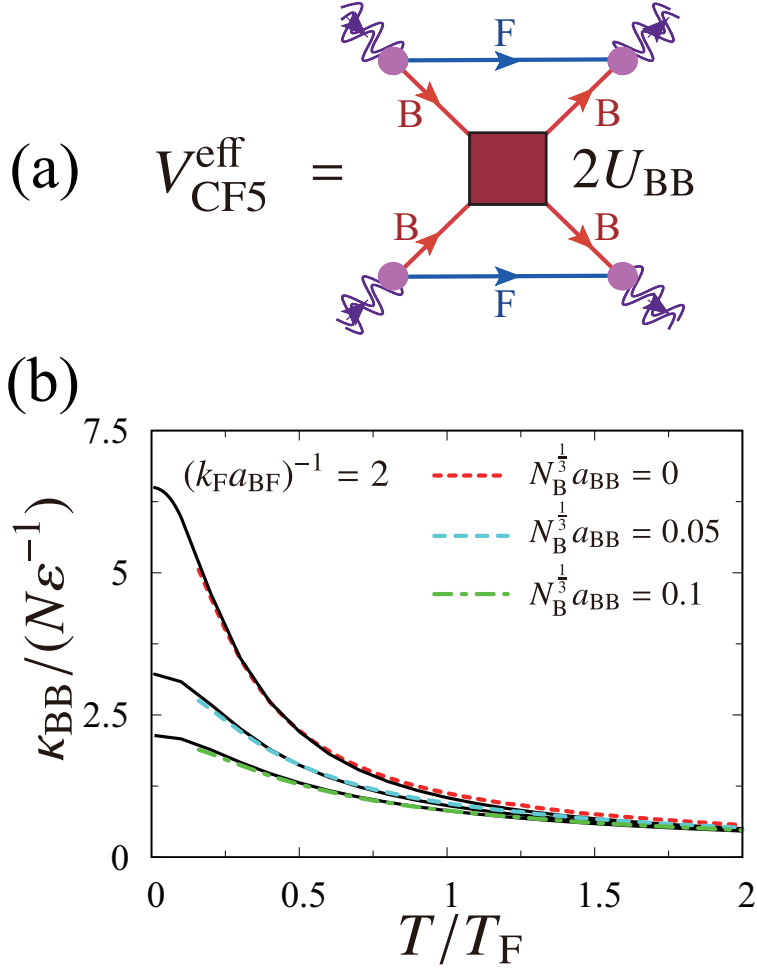}
\caption{(a) Additional effective interaction $V_{\rm CF5}^{\rm eff}$ between Fermi molecules in the presence of Bose-Bose repulsion $U_{\rm BB}$. (b) Calculated SCTMA compressibility $\kappa_{\rm BB}$ in the strong-coupling regime when $(k_{\rm F}a_{\rm BF})^{-1}=2$. The solid lines show $\kappa_{\rm CF}(U_{\rm BB}>0)$ in Eq. (\ref{kappaFFUB}). Because the other components $\kappa_{\alpha\beta}$ are almost the same, we only show $\kappa_{\rm BB}$ in this figure.}
\label{fig12}
\end{figure}
\par
In the strong-coupling regime where the system is dominated by tightly bound Bose-Fermi molecules, the direct Bose-Bose interaction $U_{\rm BB}$ brings about the additional inter-molecular interaction $V_{\rm CF5}^{\rm eff}$ in Fig. \ref{fig12}(a)\cite{kagan2004}. Evaluating this diagram as done in Sec. III.B, one finds it repulsive, having the form,
\begin{equation}
V_{\rm CF5}^{\rm eff}={8\pi a_{\rm BB} \over m}.
\label{eq.CF5}
\end{equation}
Adding this to the averaged interaction $\langle V_{\rm CF}^{\rm eff}\rangle$ in Eq. (\ref{UCFave2}), we see in Fig. \ref{fig12}(b) that the resulting $\kappa_{\rm CF}$ in Eq. (\ref{kappasc2}) agrees well with the SCTMA compressibility in the strong-coupling regime with $U_{\rm BB}>0$. Thus, the system in this regime is again found to be well described by a weakly interacting molecular Fermi gas, although the present SCTMA overestimates effects of the inter-molecular interaction on the compressibility matrix.
\par
\par
\section{Summary}
\par
To summarize, we have discussed the thermodynamic stability of a Bose-Fermi mixture in the normal state above $T_{\rm c}$. Including strong hetero-pairing fluctuations associated with a tunable Bose-Fermi attractive interaction $-U_{\rm BF}~(<0)$ within the framework of SCTMA, as well as a weak Bose-Bose repulsion $U_{\rm BB}~(>0)$ within the mean-field approximation, we calculated the compressibility matrix ${\hat \kappa}$, consisting of $\kappa_{\alpha\beta}=\partial N_\alpha/\partial\mu_\beta~(\alpha,\beta={\rm F,B})$. We then determined the collapse temperature $T_{\rm clp}$, below which the system is unstable against density fluctuations, from the weak- to strong-coupling regime in terms of the Bose-Fermi pairing interaction.
\par
When $U_{\rm BB}=0$, we showed that $T_{\rm clp}$ is always higher than the BEC phase transition temperature $T_{\rm c}$. All the matrix elements of ${\hat \kappa}$ diverge at $T_{\rm clp}$, and become negative below this temperature, indicating the simultaneous density collapse of both the Bose and Fermi components. As the origin of this instability, we pointed out that an effective Bose-Bose attractive interaction mediated by density fluctuations in the Fermi component. It makes the Bose component unstable, and this singularity is immediately brought about to the Fermi component through a Bose-Fermi coupling associated with the hetero-pairing interaction $-U_{\rm BF}$. We also clarified that this density collapse does not occur, when $(k_{\rm F}a_{\rm BF})^{-1}\gesim 1.1$. In this strong-coupling regime, most Bose and Fermi atoms form tightly bound Fermi molecules, so that the system properties are close to those of a Fermi gas. Because of this, the bosonic character of a Bose-Fermi mixture, as well as the instability associated with the induced Bose-Bose attraction, are suppressed.
\par
When $U_{\rm BB}>0$, the collapse temperature $T_{\rm clp}$ is suppressed to eventually disappear, when this repulsion is stronger than the induced Bose-Bose attraction by density fluctuations in the Fermi component. In this case, with decreasing the temperature, we can reach the BEC phase transition temperature $T_{\rm c}$, without suffering from density collapse. 
\par
However, it is still unclear whether the BEC phase is stable down to $T=0$ or the density collapse occurs at a temperature below $T_{\rm c}$. Because we have only examined the normal state in this paper, extension of the present approach to the BEC phase below $T_{\rm c}$ is an exciting future challenge. In addition, as clarified in this paper, the application of SCTMA to a Bose-Fermi mixture has room for improvement: In the weak (strong) coupling regime with respect to the hetero-pairing interaction, the calculated Fermi atomic (molecular) compressibility in SCTMA contradicts with the Pauli's exclusion principle, in the sense that it unphysically involves the contribution from the double occupancy of fermions in the same quantum state. Because this deficiency overestimates the Fermi atomic compressibility in weak-coupling regime, as well as the Fermi molecular compressibility in the strong-coupling regime, it also remains as another future problem how to overcome this problem. Since the stabilization of a Bose-Fermi mixture with a hetero-nuclear Feshbach resonance is crucial for the study of strong-coupling properties of this system, as well as for the realization of stable BEC phase, our results would contribute to the further development of this research field.
\par
\begin{acknowledgements}
We thank D. Kagamihara and R. Sato for discussions. K. M. was supported by the Keio University Doctoral Student Grant-in-Aid Program. Y.O. was supported by a Grant-in-aid for Scientific Research from MEXT and JSPS in Japan (No.JP18K11345, No.JP18H05406, and No.JP19K03689).
\end{acknowledgements}
\appendix
\section{Stability conditions for a Bose-Fermi mixture}
Thermodynamic stability of the system at fixed Bose and Fermi atomic numbers and temperature is conveniently determined from the Helmholtz free-energy functional\cite{viverit2000},
\begin{equation}
F(T,n_\alpha({\bm r}))=\int d{\bm r}
{\tilde f}(T,n_\alpha({\bm r})),
\label{helmholtz}
\end{equation}
where ${\tilde f}$ and $n_{\alpha={\rm B,F}}({\bm r})$ are the free-energy density and the density distribution in the $\alpha$ component, respectively. When we introduce small density fluctuations $\delta n_\alpha(\bm{r})=n_{\alpha}(\bm{r})-N_{\alpha}/V$ to a uniform system (with the initial density $N_\alpha/V$), and expand Eq. (\ref{helmholtz}) with respect to $\delta n_\alpha(\bm{r})$, the first-order terms are found to vanish due to the particle conservation, $\delta N_{\alpha}=\int d{\bm r}\delta n_\alpha(\bm{r})=0$. Retaining terms up to the second order, we have 
\begin{eqnarray}
\delta^2F(T,n_{\alpha}(\bm{r}))
&\equiv&
F(T,N_\alpha/V+\delta n_{\alpha}(\bm{r}))-F(T,N_\alpha/V)
\nonumber
\\
&=&
\frac{1}{2}\int d{\bm r}
\left(
\begin{array}{cc}
\delta n_{\rm F}(\bm{r}) & 
\delta n_{\rm B}(\bm{r})
\end{array}
\right)
\hat{W}
\left(
\begin{array}{cc}
\delta n_{\rm F}(\bm{r}) \\
\delta n_{\rm B}(\bm{r})
\end{array}
\right).
\label{helmholtz2nd}
\end{eqnarray}
Here,
\begin{eqnarray}
\hat{W}(T,N_\alpha/V)=
\left(
\begin{array}{cc}
\displaystyle \frac{\partial^2{\tilde f}}{\partial n_{\rm F}^2} & 
\displaystyle \frac{\partial^2{\tilde f}}{\partial n_{\rm B}\partial n_{\rm F}}
\vspace{1mm} 
\\
\displaystyle \frac{\partial^2{\tilde f}}{\partial n_{\rm F}\partial n_{\rm B}} & 
\displaystyle \frac{\partial^2{\tilde f}}{\partial n_{\rm B}^2}
\end{array}
\right)_{n_\alpha({\bm r})=N_\alpha/V}
\label{hessian}
\end{eqnarray}
is the Hessian matrix, which determines the thermodynamic stability of a Bose-Fermi mixture\cite{viverit2000,shirasaki2014,yu2011}. 
\par
The uniform system is stable, if and only if the free-energy $F(T,N_\alpha/V)$ is minimum ($\delta^2{F}<0$), that is, ${\hat W}$ must be positive definite. Using thermodynamic identities ($\mu_\alpha=\partial F/\partial N_\alpha$), Eq. (\ref{hessian}) can be written as,
\begin{eqnarray}
\hat{W}(T,N_\alpha/V)=
\left(
\begin{array}{cc}
\displaystyle \frac{\partial\mu_{\rm F}}{\partial n_{\rm F}}  & 
\displaystyle \frac{\partial\mu_{\rm F}}{\partial n_{\rm B}} 
\vspace{1mm}
\\
\displaystyle \frac{\partial\mu_{\rm B}}{\partial n_{\rm F}}  & 
\displaystyle \frac{\partial\mu_{\rm B}}{\partial n_{\rm B}}
\end{array}
\right)_{n_\alpha({\bm r})=N_\alpha/V},
\label{stability}
\end{eqnarray}
which just equals the inverse of compressibility matrix $\hat{\kappa}$ in Eq.\eqref{compressibility_matrix}. Substituting ${\hat W}={\hat \kappa}^{-1}$ into Eq. (\ref{helmholtz2nd}), we have
\begin{eqnarray}
\delta^2 F={1 \over 2}
\int d {\bm r}
\left[
{\kappa_{\rm BB} \over {\rm det}[{\hat \kappa}]}
\left[
\delta n_{\rm F}({\bm r})
-{\kappa_{\rm FB} \over \kappa_{\rm BB}}\delta n_{\rm B}({\bm r})
\right]^2
+{1 \over \kappa_{\rm BB}}
\delta n_{\rm B}({\bm r})^2
\right].
\label{definiteA}
\end{eqnarray}
Equation (\ref{definiteA}) is always positive, when $\kappa_{\rm BB}>0$ and ${\rm det}[{\hat \kappa}]>0$, that give the stability conditions in Eq. (\ref{eq.stability}). 
\par
In the same manner, Eq. (\ref{definiteA}) is always negative, when $\kappa_{\rm BB}<0$ and ${\rm det}[{\hat \kappa}]>0$. That is, when the compressibility matrix ${\hat \kappa}$ is negative definite, the system is unstable against  density fluctuations $(\delta n_{\rm F}({\bm r}),\delta n_{\rm B}({\bm r}))$.
\par
\par
\section{Estimation of $N_{\rm CF}$}
\par
As given in Eq. (\ref{Gamma2}), deep inside the strong-coupling regime ($(k_{\rm F}a_{\rm BF})^{-1}\gg 1$), the particle-particle scattering matrix $\Gamma_{\rm BF}(q)$ in Eq. (\ref{Gamma}) is reduced to the molecular Green's function. Although the simple relation in Eq. (\ref{Gamma2}) is justified only in the strong-coupling limit where the molecular dissociation no longer occurs, $\Gamma_{\rm BF}(q)$ in the strong-coupling regime still exhibits a quasi-polar structure even away from the strong-coupling limit.
Using this similarity, one can conveniently determine the molecular excitation energy $\omega_{\bm {q}}^{\rm CF}$ with momentum $\bm{q}$ from the (quasi) pole of the analytic continued particle-particle scattering matrix, within the neglect of the lifetime of molecule, as,
\begin{equation}
0 = \frac{m}{4\pi a_{\rm BF}}+{\rm Re}\left[\Pi_{\rm BF}(\bm{q},i\omega_n^{\rm F}\rightarrow \omega_{\bm{q}}^{\rm CF}+i\delta)\right]-\sum_{\bm{p}}\frac{m}{p^2}.
\end{equation}
Here, $\delta$ is an infinitesimally small positive number, and we have neglected the imaginary part of $\Pi_{\rm BF}(\bm{q},i\omega_n^{\rm F}\rightarrow \omega_{\bm{q}}^{\rm CF}+i\delta)$.
Then, simply treating the molecule as a free fermion, we estimate the number $N_{\rm CF}$ of Fermi molecules as,
\begin{equation}
    N_{\rm CF}=f_{\rm F}(\omega_{\bm{q}}^{\rm CF}).
\end{equation}
\par
\par


\begin{thebibliography}{99}
\bibitem{note_omega} In this simplified notation, while $p$ has the boson Matsubara frequency (Bose-type) in $G_{\rm B}(p)$ and $\Sigma_{\rm B}(p)$, it has the fermion Matsubara frequency (Fermi-type) in $G_{\rm F}(p)$, $\Sigma_{\rm F}$, $\Gamma_{\rm BF}(p)$, as well as $\Pi_{\rm BF}(p)$. For the $2\times 2$ matrix $\Lambda(p)$ in Eq. (\ref{G2func}), $p$ in the (11)-component is the Fermi-type and the (22)-component is the Bose-type. The types of the variables $p$ and $p'$ in ${\hat T}_0$ in Eq. (\ref{T0}) also depend on whether they appear in $G_{\rm B}$ or $G_{\rm F}$.
\bibitem{comment5} In the high temperature region where $\mu_{\rm CF}<0$, we take $k_{\rm F}^{\rm CF}=0$.
\bibitem{comment4} In computing $\langle V_{\rm CF3}^{\rm eff}\rangle$, we approximated the dressed Green's functions $G_\alpha$ to the bare ones $G_\alpha^0$. We then analytically summed up Matstubara frequencies, by expanding $V_{\rm CF3}^{\rm eff}$ with respect to the Bose and Fermi fugacities to the leading order. The resulting expression has the sum of the Bose and Fermi chemical potentials, which we evaluated by using Eq. (\ref{Gamsc}) as $\mu_{\rm F}+\mu_{\rm B}=\mu_{\rm CF}-E_{\rm b}$. Here, we determined $\mu_{\rm CF}$ from the number equation of a free molecular Fermi gas.
\end{thebibliography}
\end{document}